\documentclass{aa}
\usepackage{graphicx}
\newcommand{\chandra}{\emph{Chandra}}

\def\ltsim{\raise 2pt \hbox {$<$} \kern-1.1em \lower 4pt \hbox {$\sim$}}
\def\ltapprox{\raise 2pt \hbox {$<$} \kern-1.1em \lower 5pt \hbox {$\approx
$}}
\def\gtsim{\raise 2pt \hbox {$>$} \kern-1.1em \lower 4pt \hbox {$\sim$}}
\def\gtapprox{\raise 2pt \hbox {$>$} \kern-1.1em \lower 5pt \hbox {$\approx
$}}
\begin{document}

\title{Testing the radio halo-cluster merger scenario. The case of RXCJ\,2003.5--2323}
\author{S.~Giacintucci\inst{1,}\inst{2},
T.~Venturi\inst{2},
G.~Brunetti\inst{2},
D.~Dallacasa\inst{3,} \inst{2},
P.~Mazzotta\inst{4,}\inst{1},
R.~Cassano\inst{2},
S.~Bardelli\inst{5},
E.~Zucca\inst{5}}
\institute
{
Harvard--Smithsonian Centre for Astrophysics, 
60 Garden Street, Cambridge, MA 02138, USA
\and
INAF -- Istituto di Radioastronomia, via Gobetti 101, I-40129, Bologna, Italy 
\and
Dipartimento di Fisica, Universit\`a di Roma Tor Vergata, 
via della Ricerca Scientifica 1, I--00133, Roma, Italy
\and
Dipartimento di Astronomia, Universit\'a di Bologna, via Ranzani 1, I--40127, 
Bologna, Italy
\and
INAF--Osservatorio Astronomico di Bologna, via Ranzani 1, I--40127, Bologna, 
Italy}
\date{}
%
\titlerunning{The cluster radio halo in RXCJ\,2003.5--2323}
\authorrunning{Giacintucci et al.}
%
\abstract
{}
{We present a combined radio, X--ray and optical study of the galaxy cluster 
RXCJ\,2003.5--2323. The cluster hosts one of the largest, most powerful and 
distant giant radio halos known to date, suggesting that it  may be
undergoing a strong merger process. The aim of our multiwavelength study is to 
investigate the radio--halo cluster merger scenario.} 
{We studied the radio properties of the giant radio halo in RXCJ\,2003.5--2323
by means of new radio 
data obtained at 1.4 GHz with the Very Large Array, and at 240 MHz with the 
Giant Metrewave Radio Telescope, in combination with previously published GMRT 
data at 610 MHz. The dynamical state of the cluster was investigated by means
of X--ray $Chandra$ observations and optical ESO--NTT observations.}
{Our study confirms that RXCJ\,2003.5--2323 is an unrelaxed 
cluster. The unusual filamentary and clumpy morphology of the radio halo
could be due to a combination of the filamentary structure of the magnetic 
field and turbulence in the inital stage of a cluster merger.}
{}
\maketitle
\section{Introduction}\label{sec:intro}

RXCJ\,2003.5--2323 is an X--ray luminous and massive galaxy cluster at 
redshift z=0.317, belonging to the ROSAT--ESO Flux Limited X--ray cluster 
catalogue (REFLEX; B\"ohringer et al. \cite{boeringer04}). 
Its main properties are summarized in Tab.~\ref{tab:cluster}. 
RXCJ\,2003.5--2323 hosts a cluster scale radio source classified as giant 
radio halo (Venturi et al. \cite{venturi07}, hereinafter VGB07).
\\
Radio halos are diffuse, cluster--scale low surface brightness radio sources of
synchrotron origin found in a fraction of X--ray luminous (i.e. massive) 
clusters. 
They are not 
associated with individual galaxies but rather with the intracluster medium 
itself, and provide direct evidence for the existence of relativistic
particles and $\mu$G magnetic fields in galaxy clusters (see Ferrari et al.
\cite{ferrari08} for a recent review).
\\
The giant radio halo in RXCJ\,2003.5--2323 was discovered thanks to 
observations carried out at 610 MHz with the Giant Metrewave Radio Telescope 
(GMRT) as part of the GMRT Radio Halo Survey (VGB07 and \cite{venturi08}, 
hereinafter VGD08), devoted to a statistical study of diffuse radio emission 
in a complete sample of galaxy clusters in the redshift range z=0.2--0.4. 
With a largest linear size of $\sim$ 1.4 Mpc and a total radio power of 
logP$_{\rm 610 \, MHz}$ (W Hz$^{-1}$)= 25.53, this source is among the largest,
most powerful and most distant halos known to date. 
The radio halo in RXCJ\,2003.5--2323 has a complex morphology,
with clumps and filaments extending over hundreds of kpc, an uncommon 
feature of radio halos, which generally show a fairly regular and homogeneous 
radio brightness distribution (VGB07). 
Its peculiar radio emission is reminiscent of the giant radio halo in A\,2255 
(z=0.08), whose emission has a filamentary structure, with a typical length of 
$\sim 550$ kpc and fractional polarization at levels of $\sim 20-40 \%$ 
(Govoni et al. \cite{govoni05}). So far A\,2255 is the only known radio halo
with polarized emission.

%
%
\begin{table}\label{tab:cluster}
\caption[]{General properties of RXCJ2003.5--2323.}
\begin{center}
\footnotesize
\begin{tabular}{ll}
\hline\noalign{\smallskip}
RA$_{J2000}$  &  20h 03m 30.4s \\
DEC$_{J2000}$ & $-$23$^{\circ}$ 23$^{\prime}$ 05$^{\prime \prime}$ \\
z & 0.317  \\
L$_{\rm X \, [0.1-2.4 \, keV]}$ & 9.25 $\times 10^{44}$ erg s$^{-1}$ (a) \\
M$_{\rm V}$ &  2.05 $\times$ 10$^{15}$M$_{\odot}$ (b) \\
R$_{\rm V}$ &  2.75 Mpc (b) \\ 
\hline\smallskip
\end{tabular}
\end{center}
Notes to Table \ref{tab:cluster}: (a) B\"ohringer et al. \cite{boeringer04}; 
(b) Virial Mass M$_{\rm V}$ and virial radius R$_{\rm V}$ from the 
L$_{\rm X \, [0.1-2.4 \, keV]}$--M$_{\rm V}$ relation in Cassano, Brunetti 
\& Setti \cite{cbs06}.
\end{table}
%
%

The origin of giant radio halos in clusters has long been debated, since the
life--time of the relativistic electrons responsible for the synchrotron
radio emission is much shorter than the diffusion time necessary to cover
their Mpc extent, therefore some form of re--acceleration is invoked.
Among the possible models provided so far, the re--acceleration scenario 
(Brunetti et al. \cite{brunetti01}; Petrosian \cite{petrosian01}), which 
requires that electrons are in--situ 
re--accelerated by turbulence injected in the intracluster medium (ICM) by a 
recent or still ongoing cluster major merger event, has been recently 
given support (Feretti \cite{feretti03}; and Brunetti \cite{brunetti08} and 
Cassano \cite{cassano09} for recent reviews).

A connection between radio halos and mergers is suggested by radio
observations. Indeed, an analysis of the clusters with 
sensitive radio and X--ray imaging shows that all clusters hosting a radio
halo show signs of ongoing mergers, while those without 
may be either perturbed or relaxed (Buote \cite{buote01}; Govoni et al.
\cite{govoni04}; VGD08 and references therein).
In the case of RXCJ\,2003.5--2323, the sparse information available 
in the literature and public archives, both in the X--ray and optical bands, 
was insufficient to derive its dynamical properties (VGB07).
\\
In order to test the merger--halo connection, we carried out {\it Chandra} 
X--ray and ESO New Technology Telescope (NTT) optical observations, to 
study the dynamics of the ICM (which represents the collisional part) 
and of the cluster galaxy population (essentially collisionless).
Moreover, we observed the cluster with the GMRT at 240 MHz and with the 
Very Large Array (VLA) at 1.4 GHz, in order to derive the radio spectral
properties of the halo. 

In this paper we report on the new multi--band observations. The radio 
observations and data reduction are described in Sec.~2, and their results and 
analysis are reported in Sec.~3; the {\it Chandra} 
observations and analysis are described in Sec.~ 4; the ESO--NTT optical 
observations are presented in Sec.~5; results are presented and discussed in 
Sec.~6; summary and conclusions are given in Sec.~7.

We adopt the $\Lambda$CDM cosmology with H$_0$=70 km s$^{-1}$ Mpc$^{-1}$, 
$\Omega_m=0.3$ and $\Omega_{\Lambda}=0.7$. At the redshift of 
RXCJ\,2003.5--2323 (z=0.317) this cosmology leads to a linear scale of 
$1^{\prime \prime}=4.62$ kpc. The spectral index $\alpha$ is defined according 
to S$\propto \nu^{-\alpha}$. All the error ranges are 90\% confidence 
interval, unless stated otherwise.

\section{Radio observations and data reduction}\label{sec:obs}

In order to derive the spectrum of the giant radio halo in RXCJ\,2003.5--2323
with at least three data points and to investigate its polarization properties,
we observed the cluster at 1.4 GHz with the VLA and at 240 MHz with the  
GMRT. The details on all the observations are summarized in Tab.~\ref{tab:obs},
where we provide the observing date, frequency, total bandwidth, total time on 
source, half power bandwidth (HPBW) and rms level (1$\sigma$) in the full 
resolution images, {\it u--v} range and largest detectable structure (LDS).

%
%


%
%

%
%

\begin{table*}[t]
\caption[]{Summary of the radio observations.}
\begin{center}
\footnotesize
\begin{tabular}{lcclccccc}
\hline\noalign{\smallskip} 
Telescope & Observation & $\nu$ & $\Delta \nu$  & t  & HPBW, p.a. & rms & {\it u--v} range & LDS \\ 
          & date& (MHz)& (MHz) & (min) & (full array , 
$^{\prime \prime}\times^{\prime \prime}$, $^{\circ}$)&  ($\mu$Jy b$^{-1}$) & (k$\lambda$) & ($^{\prime}$)  \\
\noalign{\smallskip}
\hline\noalign{\smallskip}
VLA--CnB  & 9, 30 Oct 2006  & 1400  &  25 &  540  & 12.6 $\times$ 9.2, 85  & 20 & $\sim$0.27--26.5 & 7  \\
GMRT   &  3 Jun 2007 & 240 & 8 &  240 & 31.6$\times$10.8, 52 & 290 & $\sim$0.04--20.5 & 44 \\ 
\\
\noalign{\smallskip}
\hline\noalign{\smallskip}
\end{tabular}
\end{center}
\label{tab:obs} 
\end{table*}

%
%

\subsection{VLA 1.4 GHz observations}\label{sec:obsvla}

The 1.4 GHz observations were carried out in October 2006 with the VLA 
in the hybrid CnB configuration in order to optimize the u--v coverage 
at the low declination of the source. We observed the halo in two
different runs, for a total integration time of $\sim$ 9 hours on 
source (Tab.~\ref{tab:obs}). The standard polarization mode at 1.4 
GHz was used during the observations. The data were collected in 
spectral--line mode in order to better filter out the radio frequency 
interference in the observing band, and properly image the whole 
cluster field with wide--field imaging. We used 8 channels for each 
of the two IFs, centered at 1385 MHz and 1465 MHz, with a total bandwidth 
of 25 MHz/IF. 3C\,286 and 3C\,343 were observed for the bandpass and primary
flux density calibration, and for the calibration of the polarization 
electric vector. The phase calibration was obtained from the nearby 
calibrator 1923--210, while multiple observations of 1949--199 over a 
large range of parallactic angles were used to calibrate the instrumental
polarization. Calibration and imaging were performed using the 
National Radio Astronomy Observatory (NRAO) Astronomical Image Processing 
System (AIPS) package. 

The datasets from the two different days were calibrated separately. 
After bandpass calibration, the 8 channels/IF of each dataset were 
averaged to 1 single channel of $\sim$22 MHz. Each dataset was 
self--calibrated in phase only, implementing the wide--field imaging 
faceting technique to compensate for non--coplanarity.
We used a total of 
18 facets to cover the primary beam area and bright outlying sources. The 
final self--calibrated datasets were then combined together to produce 
the final images. 

Both uniform and natural weighting were used for the total intensity 
image I, and for the images in the Stokes parameters Q and U. The polarized 
intensity images were derived from the Q and U images. A high sensitivity 
(1 $\sigma=20$ $\mu$Jy b$^{-1}$) was achieved in our final full resolution 
I image (Tab.~\ref{tab:obs}). The sensitivity of the U and Q images 
is 1$\sigma$= 11 $\mu$Jy b$^{-1}$. 
The residual amplitude errors are $\ltsim$ 5\%.

\subsection{GMRT observations at 240 MHz}\label{sec:gmrt_obs}

RXCJ\,2003.5--2323 was observed using the GMRT at 240 MHz in June 
2007 for a total integration time of $\sim$4 hours (Tab.~\ref{tab:obs}). 
The observations were performed in spectral--line mode using 
the upper side band (USB) with 128 channels, with a spectral 
resolution of 62.5 kHz/channel. The total observing band is 8 MHz.
During the observations a number of antennas in the East
arm was not available due to a power outage. This led to a very 
asymmetric beam in the final images produced using the full array
(Tab.~\ref{tab:obs}). 

The data calibration and reduction was carried out using AIPS. 
The bandpass calibration was performed using the primary calibrator.
Removal of radio frequency interference (RFI) was carried out using the task 
FLGIT in AIPS, and by a subsequent careful editing of residual RFI
(see Giacintucci et al. \cite{giacintucci08} for details).
The central 84 channels were averaged to 6 channels of $\sim$0.9 MHz each after 
bandpass calibration in order to reduce the size of the dataset,
and at the same time to minimize the bandwidth smearing effects 
within the primary beam. Given the large field of view of the GMRT,
25 facets covering a $\sim 2^{\circ} \times 2^{\circ}$ field were 
used during the imaging in the self--calibration process. After a 
number of phase self--calibration cycles, the final dataset was further 
averaged from 6 channels to 1 single channel\footnote{Bandwidth 
smearing is relevant only at the outskirts of the wide field, 
and does not significantly affect the region presented and analysed 
here.}. Despite the limited number of available antennas,
we achieved a sensitivity level of 1$\sigma$=290 $\mu$Jy b$^{-1}$
in the full resolution image (Tab.~\ref{tab:obs}).
The residual amplitude errors are of the order of $\ltsim$ 5\%.

%
%
\begin{figure*}[t]
\includegraphics[angle=0,width=\hsize]{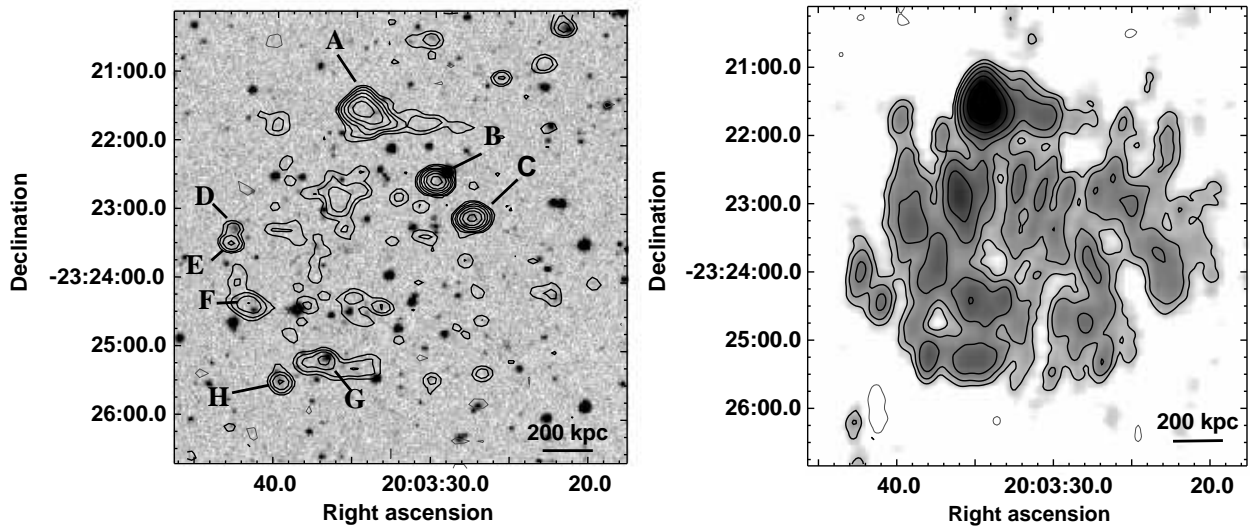}
\caption{{\it Left panel} -- Full resolution VLA 1.4 GHz contours 
of the central region of RXCJ\,2003.5--2323, superposed to the 
POSS--2 red optical image. The resolution of the radio image is 
$12.6^{\prime\prime} \times 9.2^{\prime\prime}$, 
p.a. $85^{\circ}$. The 1$\sigma$ level is 20 $\mu$Jy/b. Contours
are spaced by a factor 2, starting from $\pm 0.06$ mJy b$^{-1}$. 
Individual sources are labelled from A to H. {\it Right panel} -- VLA 1.4 GHz 
gray scale and radio contours of the giant radio halo after subtraction 
of the individual radio galaxies (from B to H in the left panel). 
The resolution is $25.8^{\prime\prime} \times 13.7^{\prime\prime}$, 
p.a. $-8^{\circ}$. The 1$\sigma$ level in the image is 20 $\rm \mu$Jy 
b$^{-1}$. Contours are spaced by a factor 2, starting from $\pm 0.06$ 
mJy b$^{-1}$.} 
\label{fig:rxcj2003_halo1.4}
\end{figure*}
%
%

\section{The giant radio halo}\label{sec:radiohalo}

\subsection{Morphology at 1.4 GHz}\label{sec:halo1400}

The VLA full resolution image at 1.4 GHz of the central region 
of RXCJ2003.5--2323 is presented in the left panel of 
Fig.~\ref{fig:rxcj2003_halo1.4}. The radio contours are overlaid 
on the POSS--2 red optical image to highlight the discrete radio 
sources in the cluster region (labelled from A to H), which were 
also detected in the $\sim 7^{\prime\prime}$ resolution image at 610 
MHz (see Fig.~6 in VGB07). All sources have an optical association, 
except source A which has no clear counterpart on the POSS--2 image.
The nature of A is unclear, as it will be briefly discussed in Sect. 
\ref{sec:genprop}.
\\
Given the high sensitivity of the image, 
the brightest regions of the diffuse emission from the central part of the 
radio halo are already clearly visible, despite the inadequate angular 
resolution. An image of the halo tapered to a resolution of 
$26^{\prime\prime} \times 14^{\prime\prime}$ is given in the right panel of 
Fig.~\ref{fig:rxcj2003_halo1.4}.  The image was obtained after subtraction 
of the optically identified radio sources (i.e. B to H).

\begin{figure}[h!]
\centering
\includegraphics[angle=0,width=8.5cm]{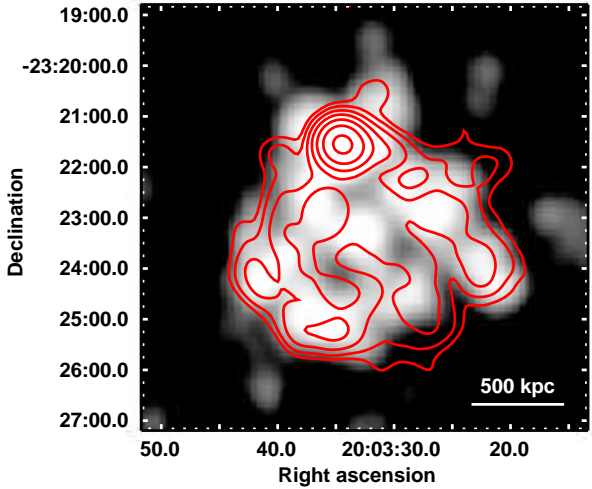}
\caption{VLA 1.4 GHz contours (red) of the giant radio halo
in RXCJ\,2003.5--2323, superposed to GMRT 610 MHz image 
(grey scale). Discrete radio galaxies have been subtracted 
in both images. The resolution is $35.0^{\prime\prime} 
\times 35.0^{\prime\prime}$, p.a. $0^{\circ}$. 
The 1$\sigma$ level is 30 $\mu$Jy b$^{-1}$ at 1.4 GHz and 
100 $\mu$Jy b$^{-1}$ at 610 MHz. Contours are spaced by a 
factor 2, starting from $\pm 0.09$ mJy b$^{-1}$}
\label{fig:rxcj2003_comp}
\end{figure}

The halo extends on a total scale of $\sim 5^{\prime}$, corresponding 
to a liner size of $\sim$ 1.4 Mpc, in very good agreement with the
source size at 610 MHz. As observed with the GMRT at 610 MHz 
(VGB07), the source exhibits a complex and inhomogeneous structure, 
characterized by 
bright clumps and filaments of emission (significant at the level of
$12\sigma$). In Fig.~\ref{fig:rxcj2003_comp} we 
compare the halo at 610 MHz (grey scale) and at 1.4 GHz (red contours). 
The images were produced using the same u--v range 
(0.2--10 k$\lambda$), and restored with the same beam of 35$^{\prime \prime} 
\times $ 35$^{\prime \prime}$. The overall size and shape of the source 
at these two frequencies are in general agreement, however the details 
of the halo surface brightness distribution, i.e. position of peaks and 
clumps, differ in the two images. 
Most likely, such discrepancy is both observational and intrinsic, i.e. partly
due to the different sensitivity and u--v coverage of the two 
observations, and partly due to a patchy distribution of the spectral
index.
%
%
%
%

\subsection{The radio halo polarization at 1.4 GHz}\label{sec:rxcj2003_pola}

The inhomogenous surface brightness distribution of the giant radio halo 
in RXCJ\,2003.5--2323 is reminiscent of the morphology of the halo in A\,2255, 
where polarized filaments of emission were detected by Govoni et al. 
(2005) with  the VLA at 1.4 GHz, with fractional polarization is of the
order of $\sim 20-40 \%$. The upper limit to the polarized emission in the 
faintest regions of the halo is of the order of $\ltsim$ $15\%$.

We used our VLA 1.4 GHz polarimetric observations (Tab.~\ref{tab:obs}) 
to search for possible polarized emission associated with the substructures 
in the surface brightness distribution of the radio halo in RXCJ\,2003.5--2323.
We obtained images of the linear polarized intensity from the full resolution 
Q and U images ($13^{\prime \prime}\times9^{\prime \prime}$), as well images 
tapered to a resolution of the order of 
$30^{\prime \prime}\times20^{\prime \prime}$.
We did not detect any significant polarized signal in the source. The upper 
limit (1$\sigma = 15\mu$Jy b$^{ħ-1}$) to the fractional polarization is 
$\sim 2-3 \%$ in the 
brightest regions of the halo (i.e. clumps and filaments) and $\sim 15\%$ 
in the faintest regions, where the average total intensity emission is about 
0.15 mJy. 
This is consistent with the polarization properties of radio halos, which are 
unpolarized, with upper limits of the order of few percent  (Govoni et
al. 2005), and confirms that A\,2255 is an exceptional case.

\subsection{Morphology at 240 MHz}\label{sec:halo235}

\begin{figure}[h!]
\centering
\includegraphics[angle=0,width=8cm]{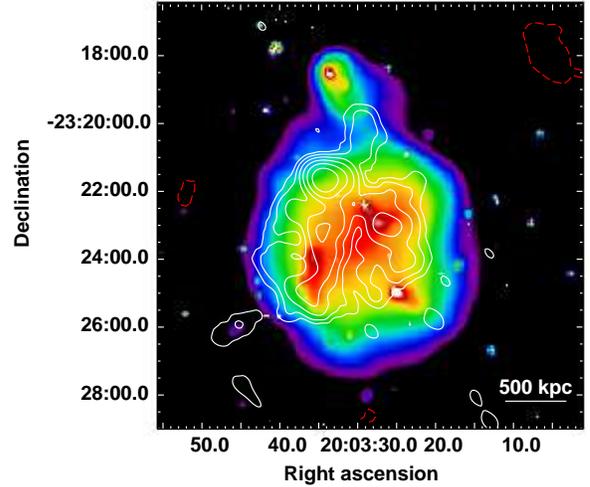}
\caption{$Chandra$--GMRT overlay. The X--ray Chandra photon image 
is background--subtracted, exposure corrected and binned to 
4$^{\prime\prime}$ pixels (see Section 4), and it includes the point
sources. The GMRT 235 MHz image (contours) of the  giant radio halo 
in RXCJ\,2003.5--2323 was obtained after subtraction of the 
individual radio galaxies. The resolution is $35.0^{\prime\prime} \times 
35.0^{\prime\prime}$, p.a. $0^{\circ}$. The 1$\sigma$ level is 
0.4 mJy b$^{-1}$. Contours are spaced by a factor 2, starting from 
$\pm 1.2$ mJy b$^{-1}$.}
\label{fig:rxcj2003_235}
\end{figure}

In Figure \ref{fig:rxcj2003_235} we present the image of radio 
halo at 240 MHz, after  subtraction of the individual radio
galaxies. The image is overlaid with the X--ray emission from
Chandra (see Sect. \ref{sec:x-ray_data} for details). 
The 240 MHz image has been convolved with a circular beam with 
HPBW=35$^{\prime \prime}$ and has a sensitivity of 1$\sigma$=400 mJy b$^{-1}$. 
The morphology of the halo at this frequency is very similar to what 
is observed at 610 MHz and 1.4 GHz, both in the total extent and in the
brightness distribution (Fig.~\ref{fig:rxcj2003_comp}).
\\
The only feature which differs in the two images is a filament of radio
emission, detected in the upper portion of the source, which departs
from source A and extends 
toward the North for $\sim1.6^{\prime}$ ($\sim 450$ kpc). Such filament, barely 
visible at 610 MHz, and almost undetected at 1.4 GHz 
(Fig.~\ref{fig:rxcj2003_comp}), has a steep spectrum, well represented 
by a single power law with $\alpha_{\rm 240~MHz}^{\rm 1.4~GHz}$=1.8.
Very little can be said about this feature at this stage, but it deserves
further investigation.

%
%

\subsection{General properties of the radio halo}\label{sec:genprop}

The overall extent and shape of the
radio halo in RXCJ\,2003.5--2323 are very similar at 240 MHz, 610 MHz and
at 1.4 GHz.
In Tab.~\ref{tab:rxcj2003} we summarize the main observational properties. 
All the given flux densities were obtained from images of comparable 
resolution 
($\sim$35$^{\prime \prime}$), and include the contribution of source A 
(Fig.~\ref{fig:rxcj2003_halo1.4}, left panel, see also 
Sec.~\ref{sec:halo1400}). In particular, the total flux density at 1.4 GHz 
was measured on the low resolution image in Fig.~\ref{fig:rxcj2003_comp}. 
The derived total radio power at 1.4 GHz is logP$_{1.4\rm \, GHz}$ 
(W Hz$^{-1}$)= 25.09. This value is consistent with the power of other 
known radio halos hosted by clusters of similar X--ray luminosity, and in
agreement with the P$_{1.4\rm \, GHz}$ -- L$_X$ correlation observed
for clusters with giant radio halos (e.g., Cassano et al. \cite{cbs06}; 
Brunetti et al. \cite{brunetti07} and VGB07).

\begin{figure}
\centering
\includegraphics[angle=0,width=8cm]{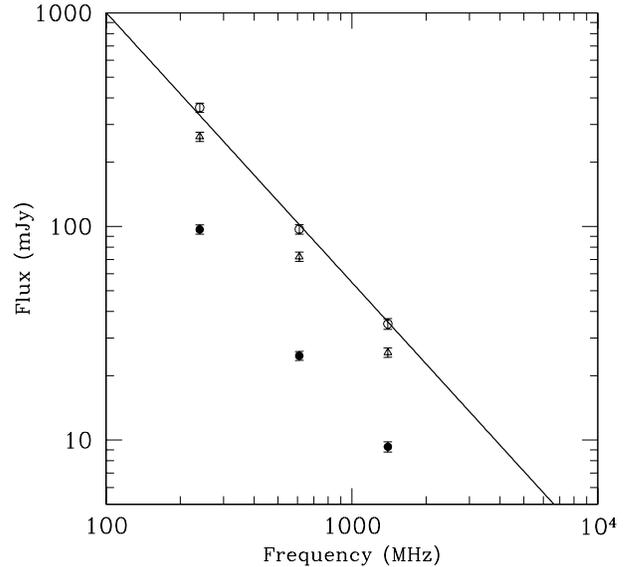}
\caption{Integrated radio spectrum of the halo between
240 MHz and 1.4 GHz. Filled circles are the data for source
A only, empty triangles are the data for the radio halo only;
empty circles are for the radio halo+source A.
The solid line is the power law fit to the data.}
\label{fig:halo_sp}
\end{figure}

The spectrum of RXCJ\,2003.5--2323 derived between 240 MHz and 1.4 GHz 
is shown in Fig.~\ref{fig:halo_sp}. It can be fitted with a single power law 
with spectral index $\alpha_{\rm fit}=1.27^{+0.18}_{-0.08}$ (solid line
in the figure), consistent with the spectral index measured in
the 240--1400 MHz range (Tab.~\ref{tab:rxcj2003}), and 
in line with the typical values 
reported in the literature for the class of giant radio halos (i.e. 
$\alpha$ $\sim$1.2--1.3; e.g. Feretti 2005). 
\\
The spectral index distribution across the radio halo did not provide 
significant results in any frequency interval, due to the substantially 
different u--v coverages, which provided a patchy spectral index image with 
random variations around the average value.

The spectrum of source A, derived with
flux density measurements obtained from the full resolution images at
each of the available radio frequencies, is consistent with that  of
the radio halo, its spectral index being $\alpha_{240~MHz}^{1.4~GHz} \sim 1.3$
(see Fig. \ref{fig:halo_sp}).
The nature of this source is unclear. It is extended in the 
East--West direction, for a total projected size of the order of $\sim$ 410 kpc,
if located at the cluster distance.
Its morphology is similar to that of cluster radio galaxies; at the same time, 
the lack of a counterpart in the optical and in the {\bf near} infra--red 
(after inspection of the SuperCOSMOS Sky Survey and of the Two Micron 
All Sky Survey respectively), as well as 
the lack of any point--like feature in the X--ray surface brightness 
(see Sect. \ref{sec:x-ray_data}) suggest
that it might be a peculiar bright spot in the radio halo.

We can give an estimate of the magnetic field associated with the
radio halo using the equipartition equations with a low energy cut--off of 
$\gamma_{\rm min}=50$ in the particle energy spectrum (Brunetti et al. 
\cite{brunetti97}), we obtain $\rm B^{\prime}_{eq}=1.7$ $\mu$G (see 
Tab.~\ref{tab:rxcj2003}).\footnote{Note that if we adopt the 
standard equipartition equations (computed in the frequency range 10 
MHz--100 GHz), we obtain $\rm B_{eq}=0.5$ $\mu$G.}
This is an average value, and should be considered as indicative of the 
cluster magnetic field.

%
%
\begin{table} 
\caption[]{Properties of the radio halo in RXCJ2003.5--2323.}
\begin{center}
\begin{tabular}{lc}
\hline\noalign{\smallskip}
S$_{1400\rm \, MHz}$ (mJy) & 35$\pm$2 \\
\noalign{\smallskip}
S$_{610\rm \, MHz}$ (mJy) & 97$\pm$5 $*$\\
\noalign{\smallskip}
S$_{240\rm \, MHz}$ (mJy) & 360$\pm$18 \\
\noalign{\smallskip}
$\alpha$ &  1.32$\pm$0.06 \\
\noalign{\smallskip}
logP$_{1400\rm \, MHz}$ (W Hz$^{-1}$) & 25.09 \\
\noalign{\smallskip}
Linear size (Mpc) & $\sim$1.4 \\
\noalign{\smallskip}
B$^{\prime}_{\rm eq}$ ($\mu$G) & 1.7  \\
\hline
\end{tabular}
\end{center}
\label{tab:rxcj2003}
\hspace{1.5cm} $*$ VGB07.
\end{table}

\section{X--ray Chandra observations and analysis}\label{sec:x-ray_data}

RXCJ2003.5--2323 was observed by {\it Chandra} on August 2 
2007 
with the ACIS‐-I detector in very faint (VF) mode. 
The details on the observation are listed in Tab.~\ref{tab:chandra}.

%
%
\begin{table}
\caption[]{Details of the {\it Chandra} observation.}
\begin{center}
\footnotesize
\begin{tabular}{cc}
\hline\noalign{\smallskip} 
Observation mode  & ACIS--I, 01236 \\
Observing date & 2007 August 2 \\
Total exposure & 50.1 ks \\
Effective exposure & 49.8 ks  \\
\hline
\end{tabular}
\end{center}
\label{tab:chandra}
\end{table}
%
%

The data were re--processed from the level 1 event files using the version 
3.4 of the Chandra Interactive Analysis of Observations (CIAO) package
and calibration database CALDB 3.4.2. The standard filtering was performed, 
excluding known bad columns, hot pixels, chip node boundaries, and events 
with ASCA grades 1, 5, and 7. We also applied the VF mode filtering, which 
significantly reduces the level of the particle background. The data were 
cleaned for flaring episodes using the recommendations given in 
\cite{2003ApJ...583...70M} leading to a useful exposure of 49.8 ksec. 
For the background subtraction we used the period D compilation of blank field 
observations provided by Markevitch\footnote{http:$//$cxc.harvard.edu$/$contrib$/$maxim$/$acisbg$/$data$/$}.
Following the prescription of \cite{2000ApJ...541..542M}, the blank-field 
dataset was first processed identically to the dataset and then reprojected 
onto the sky using the aspect information and the tool \textsc{make\_acisbg}. 
The resulting background file was finally re--normalized to take into
account the 
short--term and secular intensity variation of the changed particle background 
which dominate the spectra at high energies. 
This was done by calculating the ratio of the count rates of the observation 
to the blank-field in the energy band of $[9.5-12]~{\rm keV}$, where the 
\chandra~ effective area is nearly zero, which gives a normalization factor 
of 1.41.

\subsection{Image Analysis}

\begin{figure*}[t]
\centering
\includegraphics[angle=0,width=\hsize]{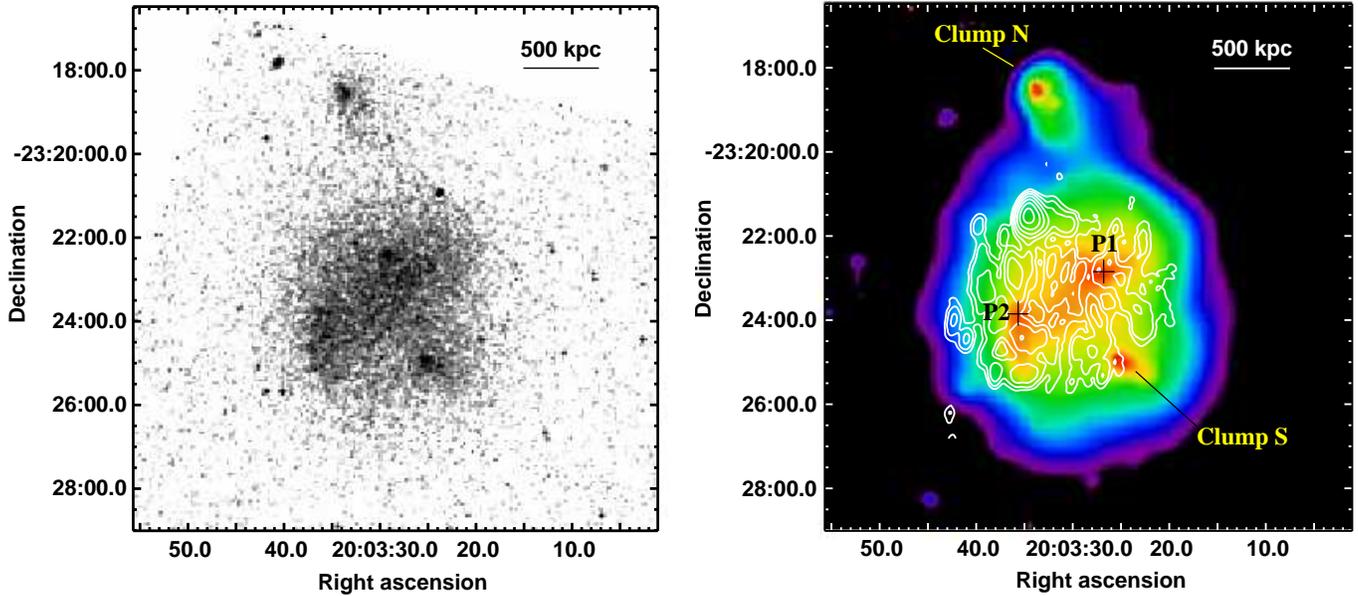}
\caption{{\it Left panel} 0.5--2.5 keV {\it Chandra} photon image of 
RXCJ\,2003.5--2323. The image is background--subtracted, exposure--corrected, 
and binned to 4$^{\prime \prime}$ pixels. {\it Right panel}
VLA 1.4 GHz contours (same as Fig.1 right panel) overlaid on the  
0.5--2.5 keV {\it Chandra} photon image smoothed with a 
gaussian with $\sigma=8^{\prime \prime}$. X--ray point sources have been
subtracted.}
\label{fig:chandra}
\end{figure*}

The {\it Chandra} photon image of RXCJ\,2003.5--2323 in the 0.5--2.5 keV band 
is presented in the left 
panel of Fig.~\ref{fig:chandra}. The image is 
background subtracted and divided by the exposure map. In the right panel 
we show the smoothed {\it Chandra} image in the 0.5-5.0 keV band 
after the subtraction of the point sources, with the radio contours at
1.4 GHz overlaid.  

The cluster X--ray surface brightness distribution is clearly complex and 
disturbed. The central part is elongated in the South--East/North--West 
direction. Two peaks of emission, labelled as P1 and P2, are separated by 
$\sim$ 2.5$^{\prime}$ ($\sim$ 700 kpc) along this axis. A bright compact 
clump is located in the northern region (clump N), at a projected distance 
of $\sim$1.2 Mpc from P1. A second sub--structure (clump S) is detected at 
$\sim$600 kpc South of P1. 

Except for an overall agreement in the shape of the radio and of the X--ray 
emission, there is very little coincidence between the radio and the X--ray 
images: the S--E/N--W elongation of the inner X--ray brightness 
distribution, has no corresponding features in the radio halo emission, 
and the two peaks P1 and P2 have no association with discrete radio sources.
Moreover, the superposition between the radio and the X--ray emission
indicates that the two clumps N and S have no radio counterpart. The only
possible hint of connection with clump N is given in Fig. 
\ref{fig:rxcj2003_235}, which shows that the radio halo emission at 240 MHz
is characterized by a steep spectrum filament extending from source A 
towards clump N.

\subsection{Spectral analysis}

The spectral analysis was carried out using 
the version 12.2 
of the X--Ray Spectral Fitting Package (XSPEC). After the detection and 
removal of the point sources within the cluster emission, we extracted the 
spectra from the selected regions of both the observation and the background.
We then created the relative Redistribution Matrix File (RMF) and Auxiliary 
Response File (ARF). The spectra were grouped in order to have at least 20 
counts per channel, and fitted using XSPEC with an absorbed single temperature 
thermal model ({\textsc WABS*MEKAL}). The fit was performed in the energy 
range $[0.6- 8]$~keV. 


We determined the global properties of the cluster by extracting the spectrum 
from a circular region with radius r=4$^{\prime}$ (i.e. $\sim$ 1.1 Mpc) 
centered on the surface brightness centroid (RA$_{\rm J2000}$= 20h 03m 29s, 
DEC$_{\rm J2000}$=$-$23$^{\circ}$ 23$^{\prime}$ 36$^{\prime \prime}$). 
The fit was made by fixing the redshift to z=0.317 and the hydrogen absorption 
column to the galactic value ($\rm N_H=8.57\times 10^{20}$ cm$^{-2}$) and 
letting the chemical abundance $Z$ and the temperature $T$  free to vary. 
We found T=9.1$^{+0.7}_{-0.6}$~keV and  Z=0.27$\pm$0.10. We also 
let $\rm N_H$ free to vary and found that the fitted value of 
$\rm N_H=(9.30 \pm 1.73) \times 10^{20}$ cm$^{-2}$ is perfectly consistent 
with the galactic value. The temperature found for this cluster is
consistent with the values known in the literature for clusters 
hosting giant radio halos (Cassano et al. \cite{cbs06}).

%
%
%
%

\subsection{Hardness Ratio Image}\label{sec:hratio}

\begin{figure}[t]
\centering
\includegraphics[angle=0,width=\hsize]{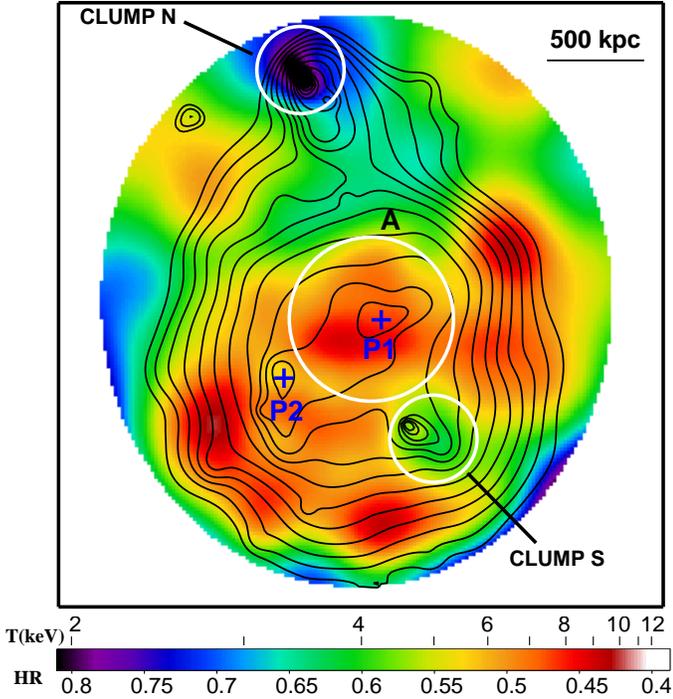}
\caption{Hardness ratio image of RXCJ\,2003.5--2323. The smoothed image 
of the cluster in the 0.5--2.5 keV band is overlaid as contours, 
logarithmically spaced by a factor $\sqrt(2)$ (point sources 
were removed). The regions used for the spectral fitting, and the
temperature calibration are reported.}
\label{fig:hratio}
\end{figure}

Given the complex structure of the X--ray surface brightness
distribution of the cluster (Fig.~\ref{fig:chandra}), we searched 
for possible substructures in the gas temperature distribution.
We produced the hardness ratio image of the cluster using the two 
photon images in the 0.2--2.5~keV ({\it soft}) and 2.5--7.0~keV ({\it hard}) 
energy bands, both background subtracted and vignetting corrected. 
After removing the point sources, the {\it soft} and {\it hard} images 
were adaptively smoothed, using the same smoothing scale pattern (map).
The smoothing scale map was obtained by applying the CIAO command 
{\it csmooth} to the {\it soft} image requesting a minimum and a maximum 
significance of the signal under the smoothing kernel of $7\sigma$ and 
$10\sigma$, respectively. The resulting smoothed images were combined 
as ({\it soft} -- {\it hard})/({\it soft} + {\it hard}) to obtain the 
hardness ratio image shown in Fig.~\ref{fig:hratio}. The contours are 
obtained from the smoothed {\it soft} image. Finally we 
converted the hardness ratio values in the corresponding gas temperatures 
by simulating a number of absorbed single temperature thermal models with 
XSPEC. The result of the conversion is reported on the color bar of 
Fig.~\ref{fig:hratio}.

%
%
%
%

%
%
\begin{figure}[h!]
\centering
\includegraphics[angle=0,width=8cm]{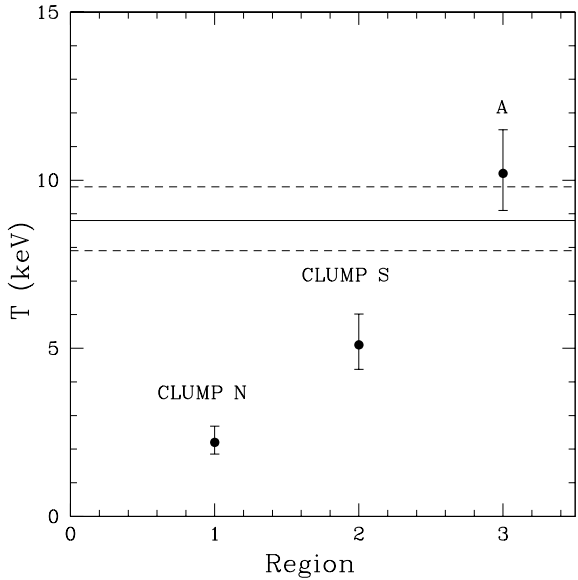}
\caption{Best--fit temperature of the regions defined in the right 
panel of Fig.~\ref{fig:hratio}. The solid line is the average temperature of 
the cluster (Tab.~\ref{tab:temp2}); the dashed lines delimit its error range 
at 90\% confidence interval.}
\label{fig:temp}
\end{figure}

\begin{figure}
\centering
\includegraphics[angle=0,width=9cm]{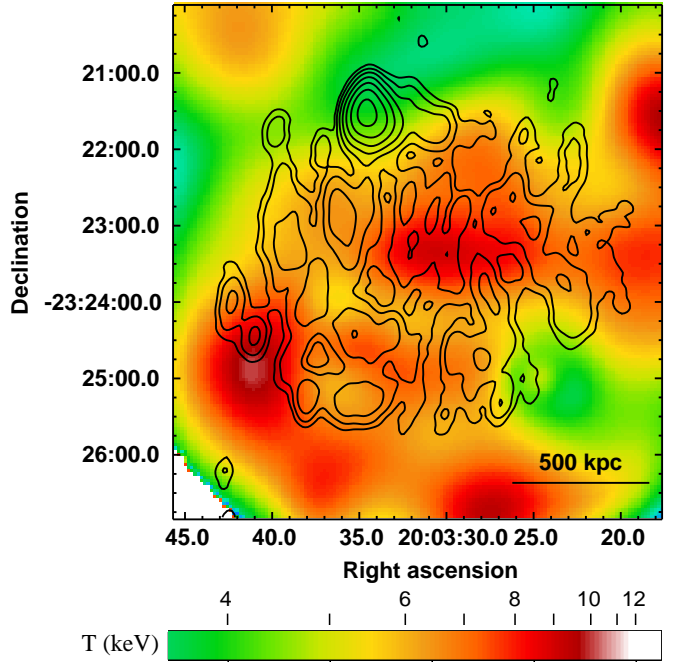}
\caption{Radio halo contours at 1.4 GHz (same as right panel of 
Fig.~\ref{fig:rxcj2003_halo1.4}) on the hardness radio image of 
RXCJ\,2003.5--2323 (same as Fig.~\ref{fig:hratio}).}
\label{fig:hratio_radio}
\end{figure}

The hardness ratio map clearly shows that  RXCJ\,2003.5--2323 has a 
complex thermal structure: it lacks a central cool core, and presents
a number of hot clumps.
The brightest innermost cluster region  is hot, with $<$T$>$ $\sim$ 8--9 keV. 
The two X--ray peaks P1 and P2 have slightly different temperatures: 
the western peak P1 has $T\approx ~8~{\rm keV}$ while the eastern peak P2 has 
$T\approx ~6~{\rm keV}$. 
Clump N, instead, is much colder ($~\ltsim$ 3 keV) than the average temperature 
of the cluster, while  clump S has a temperature of $\sim 5$ keV. 
A number of other peaks are visible in Fig.~\ref{fig:hratio}, however due 
to the low number of counts they are less significant.

To test the significance of the substructure observed in the hardness 
ratio image, we also extracted the spectra from the white circles shown 
in Fig.~\ref{fig:hratio}. The results, summarized in Table \ref{tab:temp2} 
and shown in Fig.~\ref{fig:temp}, clearly show that the detected thermal 
structures are highly significant, confirming the unrelaxed status
of this cluster. We point out that the temperatures given for P1 and P2 
should be considered as estimates, which we derived from the hardness ratio 
image, since the statistics here is too low to fit them as separate regions.

As a further comparison in our analysis, in Fig. \ref{fig:hratio_radio}
we overlaid the radio halo contours at 1.4 GHz to the hardness ratio image,
and concluded that the two images are overall independent.
%
%
\begin{table}
\caption[]{Temperatures of substructures}
\begin{center}
\footnotesize
\begin{tabular}{cccc}
\hline\noalign{\smallskip}
Region & T &  $\chi_{\rm red}^2$ & Null--hyp.\\
       & (keV) &                   &  prob. \\
\hline\noalign{\smallskip}
Clump N &  2.2$^{+0.5}_{-0.3}$ & 0.72 & 0.784 \\
Clump S &  5.1$^{+0.9}_{-0.7}$ & 0.69 & 0.747 \\
Region A & 10.2.$^{+1.3}_{-1.1}$ &  0.99 & 0.544 \\
\hline\noalign{\smallskip}
\end{tabular}
\end{center}
\label{tab:temp2}
\end{table}
%

\section{Optical observations}

We  observed the cluster in the night of 06/04/2007 with 
the EMMI instrument of the New Technology Telescope (NTT) at the ESO La Silla 
Observatory (Chile). The exposure time was 
300 seconds with the R Johnson-Cousin filter. The image was analysed with 
the standard bias subtraction and flat--fielding, and the SEXTRACTOR code 
(Bertin \& Arnouts \cite{bertin96}) was used to extract a source catalogue. 
The magnitude calibration was done on the basis of the reference stars present
in the field. We extracted 1059 objects with the SEXTRACTOR parameter 
"classstar" $<$ 0.2 (to eliminate star contamination) and with $R<23.3$ 
(value at which the magnitude histogram starts to decrease). 

We studied the substructure by means of the DEDICA algorithm (Bardelli et al. 
\cite{bardelli98}; Pisani \cite{pisani96}), and restricted the analysis to 
the projected two dimensional distribution of the galaxies.
The presence of background (i.e. non--cluster) galaxies is 
represented as a flat distribution within the image, therefore they do not 
influence the grouping algorithm and can be considered as an additional 
contribution to the noise. In order to increase the signal, we weighted all 
galaxies with their magnitude, if $18.3<R<23.3$. This approach allows
to emphasize the presence of the most likely cluster members (the brightest 
central objects) with respect to the fainter background ones. 
In practice, assuming an approximately constant mass--to--light
ratio, our light weigthed procedure provides a better estimate of the mass
distribution.

\begin{figure}
\centering
\includegraphics[angle=0,width=\hsize]{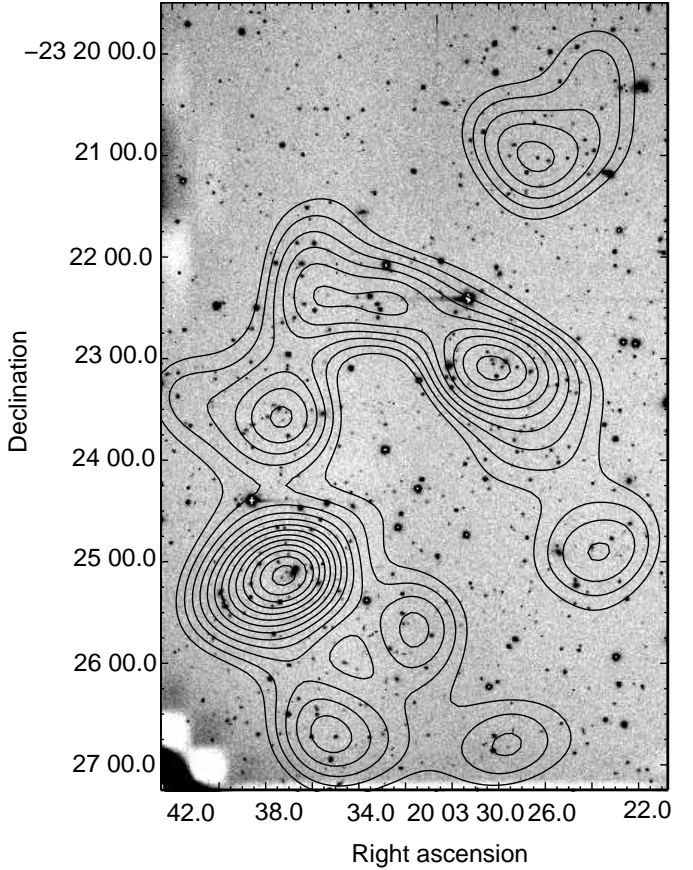}
\caption{Luminosity weighted isodensity contours superimposed to our NTT 
R image of RXCJ\,2003.5--2323}
\label{fig:ottico1}
\end{figure}

In Fig. \ref{fig:ottico1} we show the luminosity weighted isodensity 
contours superimposed to our R image. All peaks defined by more than three 
isodensity contours are significant at the $99.5 \%$ level. The richest 
condensation 
is located at RA=$20^h 03^m37^s$ and DEC=$-23^o 25^{\prime}08^{\prime\prime}$ 
and is centered on a galaxy, which could be considered the Brightest Cluster 
Galaxy. 
The second one is located at RA=$20^h 03^m 28^s$ and 
DEC=$-23^o 23^{\prime}05^{\prime\prime}$, at $\sim 3^{\prime}$ (corresponding 
to $\sim 0.8$ Mpc) from the main one. Considering a circle of 0.8 arcmin 
around the centres of these two condensations, the enclosed luminosty density 
of the Western group is $\sim 80$\% that of the Eastern one. 

\begin{figure*}[t]
\centering
\includegraphics[angle=0,width=\hsize]{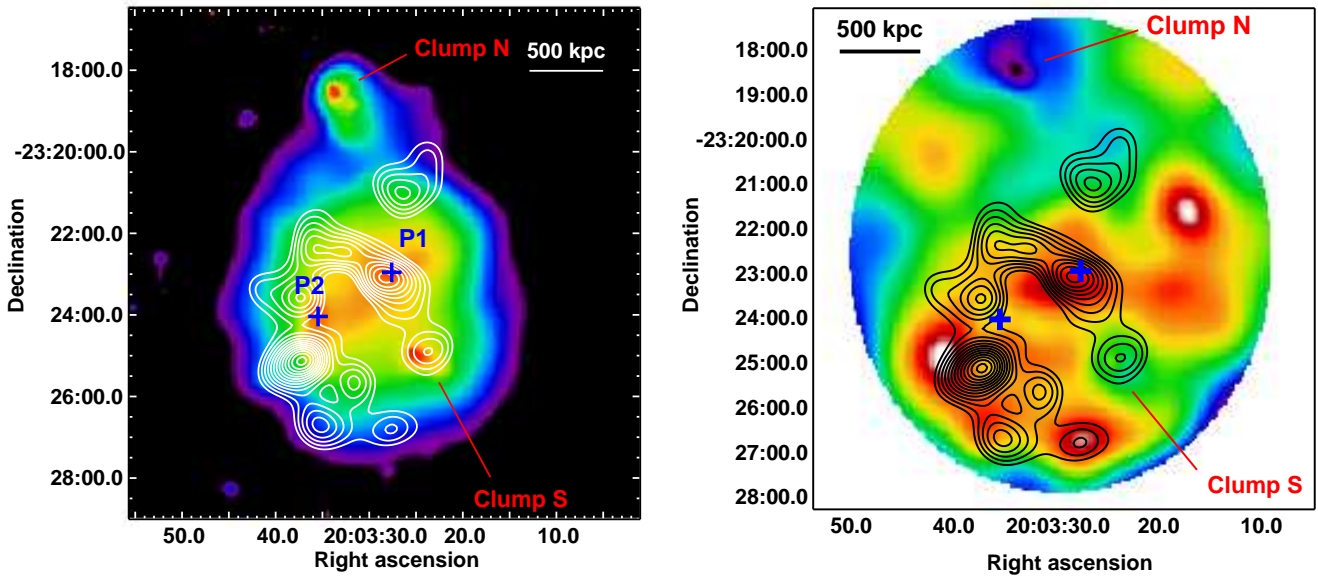}
\caption{Luminosity--weighted isodensity contours of the optical 
galaxy distribution, overlaid on the 0.5--2.5 keV smoothed {\it Chandra} 
image (left panel; X--ray point sources have been subtracted) and hardness 
ratio image of RXCJ\,2003.5--2323 (right panel; same as Fig.~\ref{fig:hratio}.}
\label{fig:chandra_opt}
\end{figure*}

Inspection of the overlay between the X--ray surface brightness 
distribution and the optical luminosity weighted contours, given in the left 
panel of Fig. \ref{fig:chandra_opt}, is very insightful. The central 
X--ray elongation and the two main optical condensations are aligned 
along the same axis. The X--ray peak P1 is coincident with the Western optical 
group, while there is an offset between P2 and the Eastern most massive 
optical condensation.
\\
An optical group is  found in coincidence with clump S, while
no optical overdensity was found in coincidence with the X--ray clump N, as 
clear from the left panel of Fig. \ref{fig:chandra_opt}.
The right panel of Fig. \ref{fig:chandra_opt} shows the luminosity-weighted
isodensity contours overlaid on the hardness ratio image. The two main
optical condensations show an offset with respect to the peaks in the hardness 
ratio.

The comparison between the optical image and the radio halo shows that
no features in the radio halo are detected in coincidence with the main 
optical condensations. 

The redshift of this cluster was estimated by  B\"ohringer et al. (2004) 
on the basis of four galaxies and turned out to be z=0.3171. During our 
observing run, we obtained the spectra for 5 more
cluster galaxies: the mean redshift of RXCJ\,2003.5--2323 resulted  to be 
$<z> = 0.314 \pm 0.003$, consistent with the published 
value assuming a similar error for the literature data.

\section{Discussion}

Our multifrequency study of the cluster of galaxies RXCJ\,2003.5--2323
confirms that it is in a complex dynamical state, and provides
important pieces of information for our understanding of the origin
of its giant radio halo, whose discovery was reported in VGB07.
The main observational results derived from our radio analysis can
be summarized as follows:

\begin{itemize}
\item[(1)] the high sensitivity GMRT and VLA radio observations confirm 
that it is one of the largest, most powerful and most distant giant radio 
halos known so far;

\item[(2)] the radio halo is characterized by a very irregular
surface brightness distribution, with clumps and filaments;

\item[(3)] the synchrotron radio spectrum is well fitted by a single
power law with a spectral index 
$\alpha_{\rm 240~MHz}^{\rm 1.4~GHz} = 1.27^{+0.18}_{-0.08}$, consistent with the 
typical values of the radio halos known in the literature;

\item[(4)] it is unpolarized, with an upper limit of the order of 
$\sim$ 2--3 \%.

\end{itemize}

In the following sections we will discuss the multiband properties 
of RXCJ\,2003.5--2323 and the possible origin of its radio halo in the 
framework of the re--acceleration scenario.

\subsection{Dynamical state of the cluster and the radio halo/cluster merger scenario}

Radio halos are always found in merging galaxy clusters, and this observational
evidence provides strong constraints to the theoretical models for the
formation of such structures. 
The X--ray and optical analysis based on our observations suggests that 
also RXCJ\,2003.5--2323 hosts a cluster merging event.
In particular the main X--ray observational properties are:

\begin{itemize}

\item[--] the X--ray brightness distribution shows substructure, as typical 
in dynamically unrelaxed clusters. It is elongated in the NW--SE
direction, with two peaks, P1 and P2, aligned in this direction, and 
separated by $\sim$ 0.8 Mpc. Two more clumps of emission, N and S, have been 
identified in the outer regions of X--ray emission, aligned in a direction 
almost orthogonal to that of P1 and P2;

\item[--] the cluster X--ray temperature, T=9.1$^{+0.7}_{-0.6}$ keV, is 
consistent with what is found in clusters hosting giant radio halos;

\item[--] temperature gradients were detected in the X--ray gas distribution.
The central region, peaked in P1, has a temperature as high as 10.2 keV,
while the temperature in the two clumps N and S drops to 2.2 and 5.1 keV
respectively. This result is further strengthened by the complex thermal 
structure revealed by the hardness ratio analysis.

\end{itemize}

The optical analysis provides the following results:

\begin{itemize}

\item[--] our spectral data confirm that the cluster is located at a
redshift $<z>=0.314 \pm 0.003$;

\item[--] the galaxy luminosity weighted distribution shows two
condensations, aligned along the NW--SE direction. The NW peak is located
in coincidence with peak P1 in the X--ray surface brightness distribution,
while the SE peak, which is also the most massive, is slighlty offset 
with respect to P2, though aligned along the same direction;

\item[--] a filament of galaxies connects the two optical peaks. 
An optical condensation was found in concidence with clump S, while
we did not find any optical overdensity associated with clump N, based
on our single band image.

\end{itemize}

Comparison of the optical and X--ray properties is also suggestive of 
dynamical activity. In particular, the shift between the location of the 
X--ray peak P2 and the most massive optical condensation, clearly visible 
in the left panel of Fig. 10, is intriguing, as well as the relative shifts 
among P2, the optical clump and the peak in the hardness ratio image. 
We thus conclude that the overall optical and X--ray properties of
RXCJ\,2003.5--2323 lead to a plausible scenario where a major merging is 
taking place in this cluster.
\\
An optical/X--ray shift
has been observed in the bullet cluster (1E\, 0657--56, Markevitch et al. 
\cite{markevitch04} and references therein) and in MACS\,J0025.4--1222
(Brada\u{c} et al. \cite{bradac08}), and has been interpreted as the result
of a galaxy sub--cluster exiting the core of the main cluster, just ahead
of the gas. In the case of RXCJ\,2003.5--2323 it is possible that 
the main optical condensation  is just emerging South--East after the 
merger, leaving the gas behind.

\subsection{Origin of the radio halo morphology}
\label{sec:rxcj2003_origin}

Giant radio halos typically exhibit a fairly regular and homogeneous 
morphology, with a flux density  distribution usually peaked at the cluster 
centre, and smoothly decreasing towards the cluster periphery
(e.g. A\,2163; Feretti et al. \cite{feretti01}). The shape and extent of the 
radio halo in RXCJ\,2003.5--2323 is in agreement with the X--ray surface 
brightness emission, however its radio flux density distribution shows clumps 
and filaments at all frequencies 
(Figs.~\ref{fig:rxcj2003_halo1.4},~\ref{fig:rxcj2003_comp} and
\ref{fig:rxcj2003_235}). 
This morphology poses the question of the connection with the ICM distribution. 

In the framework of the re--acceleration model 
powerful giant radio halos at $z > 0.3$ should  be hosted by massive, 
highly turbulent and dynamically disturbed clusters (Cassano, Brunetti 
\& Setti 2006).

The clumpy radio emission in RXCJ\,2003.5--2323  may be driven both by the 
distribution of the emitting relativistic particles, and by that of the 
magnetic field intensity in the ICM. 
\\
RXCJ\,2003.5--2323 is at relatively high redshift, where
Inverse Compton losses are strong and acceleration could produce 
synchrotron emission preferentially in regions with $B\sim B_{\rm cmb}$,
where $B_{cmb}$ is defined as $B_{cmb} \sim 3 (1+z)^2$ $\mu$G. As a matter of 
fact, synchrotron emission from lower $B$ would require the
acceleration of electrons with significantly lower energy, while larger $B$ 
would increase synchrotron losses making the acceleration process less
efficient. 
At the same time, RXCJ\,2003.5--2323 could be in an early merging phase and 
both the magnetic field and turbulence may have a filamentary structure,
which are traced by the synchrotron radiation.

Another interpretation of the origin of the Mpc scale diffuse emission
in RXCJ\,2003.5--2323 is that it is a relic seen in face--on projection. 
However, this possibility seems very unlikely, due to the lack of
polarization and to the overall spatial connection between the radio emission, 
the X--ray surface brightness and the galaxy distribution.

\section{Summary and conclusions}
\label{sec:rxcj2003_summary}

In this paper we presented a multifrequency study of the galaxy cluster 
RXCJ\,2003.5--2323, which hosts one of the most distant, largest and most 
powerful radio halos known to date, with the aim of testing the cluster 
merger--radio halo connection.

The radio halo was imaged and analysed at high sensitivity 
using the VLA at 1.4 GHz, in combination with new GMRT data at 240 MHz and 
previously published GMRT data at 610 MHz (VGB07). The halo extends on a
scale of the order of $\sim$1.4 Mpc at all frequencies,  and its overall radio
morphology is very similar in all images. The most striking feature of this 
source is its complex and uncommon radio brightness distribution, with 
clumps and filaments of emission extending on scales of the order of
hundreds of kpc, clearly visible at all radio frequencies. 
This radio morphology is very different from what is  found in giant 
radio halos, whose radio flux density generally peaks at the cluster 
centre and smoothly decreases towards the outskirts. 

Our X--ray and optical analysis suggest that the cluster is in a merging stage. 
In particular, our {\it Chandra} observations show that the cluster is very
unrelaxed, with substructure in the X--ray surface brigthness  distribution, 
in the temperature and hardness ratio. An optical analysis based on ESO--NTT 
observations shows that the galaxy distribution is characterized by two peaks
connected by a filament of galaxies, the South--Eastern peak considerably more 
massive than the North--Western one. The galaxy distribution is in
reasonable agreement with the substructure in the X--ray surface brightness
emission and in the hardness ratio image. 
{A misplacement between the X--ray peak P2 and the main optical 
condensation suggests that the latter might be emerging South--East after 
the merger event, leaving the gas behind.

The clumpy and filamentary morphology of the radio halo was discussed in the 
framework of the re--acceleration model. Due to the relatively high redshift
of the cluster, which could be in an early merging phase, it is likely that 
the observed clumps and  filaments trace the peaks of the magnetic field 
intensity and turbulence in the cluster. 
\\
\\
{\it Acknowledgements.}
We thank the staff of the GMRT for their help during the observations.
The GMRT is run by the National Centre for Radio Astrophysics of the Tata 
Institute of Fundamental Research. The optical data were obtained with the
European Southern Observatory NTT telescope, La Silla, Chile, program
079.A--0191(A). We acknowledge financial contribution 
from the Italian Ministry of Foreign Affairs, from MIUR grants PRIN2004,
PRIN2005 and 2006, from PRIN--INAF2005 and from contract ASI--INAF I/023/05/01.

\end{document}